\documentclass[preprintnumbers,amsmath,amssymb]{revtex4}

\usepackage{epstopdf}
\usepackage{graphicx}
\usepackage{dcolumn}
\usepackage{bm}
\usepackage{amsmath}
\usepackage{hyperref}

\newcommand{\be}{\begin{equation}}
\newcommand{\ee}{\end{equation}}
\newcommand{\beq}{\begin{eqnarray}}
\newcommand{\eeq}{\end{eqnarray}}

\def\t13{\mathrel{{\theta_{13}}}}
\def\y12{\mathrel{{\tan^2 \theta_{12}}}}
\def\c2{\mathrel{{\chi^2 }}}

\def\lp{\mathrel{{{l^\prime}}}}
\def\jp{\mathrel{{{j^\prime}}}}
\def\np{\mathrel{{{n^\prime}}}}

\def\qx{\mathrel{{q \mathbf{ x}}}}
\def\bm{\mathrel{{\mathbf{M}}}}
\def\by{\mathrel{{\mathbf{Y}}}}
\def\be{\mathrel{{\mathbf{e}}}}
\def\bnab{\mathrel{{ \mathbf{\nabla}}}}
\def\bsig{\mathrel{{ \mathbf{\sigma}}}}

\newcommand{\mt}{Mathematica}

\begin{document}

\preprint{INT PUB 07-15}

\title{A Mathematica script for harmonic oscillator nuclear matrix elements arising in semileptonic electroweak interactions}

\author{Wick Haxton}
 \email{haxton@u.washington.edu}   
\affiliation{Institute for Nuclear Theory and Department of Physics, University of Washington, Seattle, WA 98195 }%
\author{Cecilia Lunardini}
 \email{lunardi@u.washington.edu}   
\affiliation{Institute for Nuclear Theory and Department of Physics, University of Washington, Seattle, WA 98195 }%


 
\begin{abstract}
Semi-leptonic electroweak interactions in nuclei -- such as $\beta$
decay, $\mu$ capture, charged- and neutral-current neutrino reactions,
and electron scattering -- are described by a set of multipole
operators carrying definite parity and angular momentum, obtained by
projection from the underlying nuclear charge and three-current
operators.  If these nuclear operators are approximated by their
one-body forms and expanded in the nucleon velocity through order
$|\vec{p}|/M$, where $\vec{p}$ and $M$ are the nucleon momentum and
mass, a set of seven multipole operators is obtained.
Nuclear structure calculations are often performed in a basis of
Slater determinants formed from harmonic oscillator orbitals, a choice
that allows translational invariance to be preserved.
Harmonic-oscillator single-particle matrix elements of the multipole
operators can be evaluated analytically and expressed in terms of
finite polynomials in $q^2$, where $q$ is the magnitude of the
three-momentum transfer.  While results for such matrix elements are
available in tabular form, with certain restriction on quantum
numbers, the task of determining the analytic form of a response
function can still be quite tedious, requiring the folding of the
tabulated matrix elements with the nuclear density matrix, and
subsequent algebra to evaluate products of operators.  Here we provide
a Mathematica script for generating these matrix elements, which will
allow users to carry out all such calculations by symbolic
manipulation.  This will eliminate the errors that may accompany hand
calculations and speed the calculation of electroweak nuclear cross
sections and rates.  We illustrate the use of the new script by
calculating the cross sections for charged- and neutral-current
neutrino scattering in $^{12}$C.
\end{abstract}

\maketitle

\section{Introduction}
\label{intro}
A common task in nuclear physics is to connect an observable -- a rate for $\beta$ decay, the diffraction pattern seen in inelastic electron scattering, etc. -- back to the underlying nuclear
structure physics.  While this can always be done implicitly through numerical calculations, one of the attractive properties of the harmonic-oscillator shell model is that this connection can be made analytically, at least in the case of one-body observables \cite{walecka}. That is, each observable can be expressed as a simple function involving finite polynomials in $q^2$, the square of the three-momentum transfer to the nucleus, with the polynomial coefficients depending on the one-body density matrix.
This allows one to determine quickly how a given experimental result constrains the underlying
nuclear physics (the one-body density matrix), and to dentify the 
unconstrained degrees of freedom in the density matrix that will enter into predictions of other observables, and thus will have to be taken from models.

Specifically, electroweak processes such as $\beta$ decay, muon capture, charged- and neutral-current neutrino reactions, electron scattering, and photo-absorption and $\gamma$ decay are traditionally described as one-body processes, possibly with corrections added to account for exchange currents, final- or initial-state distortions of the lepton wave function, etc.  That is, the charge and three-current
nuclear operators are expressed as the sum of the currents of the constituent nucleons.  As described below, the many-body matrix elements of such operators can then be expressed as
sums over single-nucleon matrix elements.   Some time ago harmonic oscillator matrix elements -- that is, the coefficients of the polynomials mentioned above -- were tabulated by Donnelly and Haxton
\cite{donnelly}, though with some restrictions to reduce the size of the compilation.   The tables
include single-nucleon states up to the closed shell at proton/neutron number 126, e.g.,
all oscillators shells for principal quantum number $N$ = 0 to 5, as well as the $j=13/2$
subshell for $N$=6.  ($N$ determines the harmonic oscillator energy,
$(N+3/2)\hbar \omega$.)  The tables also restricted transitions to $|N_f-N_i| \le 2$. 

In today's context, the main shortcoming of the tables is not these restrictions, but the need
to transcribe the results when using them in calculations.  This is tedious and provides opportunities
for errors.  Thus here we provide a Mathematica \cite{wolfram} script to generate equivalent results, without
restrictions on the choice of quantum numbers.   This script will allow users
to efficiently manipulate results within Mathematica, so that
cross sections, rates, and relationships between various observables can be obtained
quickly and without error, in closed form.    

\section{One-body Operators and the Density Matrices}
We consider electroweak  nuclear multipole operators $\hat{O}_{J;T}$ that are one-body, e.g.,
have the first-quantized form $\sum_{i=1}^A O_{J;T}(i)$, and carry definite parity and angular momentum.  This operators, which are formed from the single-nucleon currents and take into account the spatial extent of these currents, can then be evaluated in terms of the one-body density matrix,
\begin{equation}
\langle J_f ; T_f \vdots \vdots \hat{O}_{J;T} \vdots \vdots J_i ; T_i \rangle =
\sum_{|\alpha|,|\beta|}
\psi_{J T}^{f,i}(|\alpha|,|\beta|) \langle |\alpha|; {\textstyle{1 \over 2}} \vdots \vdots  O_{J;T} \vdots \vdots |\beta|;{\textstyle {1 \over 2}} \rangle.
\label{eq1}
\end{equation}
Here $(J;T)$, $(J_f;T_f)$ and $(J_i;T_i)$ are the angular momentum and isospin of the operator and of the final and initial nuclear states, respectively, 
$|\alpha| = \{n_\alpha (l_\alpha {\textstyle {1 \over 2}}) j_\alpha \}$ represents the
set of nonmagnetic single-nucleon spatial and spin quantum numbers, and
$\vdots \vdots$ indicates a matrix element reduced in both angular momentum and isospin, i.e.,
\begin{eqnarray}
&&\langle J_f M_f; T_f M_{T_f}  | \hat{O}_{J;T} | J_i M_i; T_i M_{T_i} \rangle = \nonumber \\
&&(-1)^{J_f-M_f + T_f-M_{T_f}}
 \left( \begin{array}{ccc} J_f & J & J_i \\ -M_f & M & M_i \end{array} \right)
\left( \begin{array}{ccc}  T_f & T & T_i \\ -M_{T_f} & M_T & M_{T_i} \end{array} \right)
\langle J_f ; T_f   \vdots \vdots \hat{O}_{J;T} \vdots \vdots J_i ; T_i  \rangle.
\end{eqnarray}
The sums extend
over complete sets of quantum numbers $|\alpha|$ and $|\beta|$ -- which would normally
correspond to the
single-nucleon basis used in constructing the Slater determinants for the initial and final nuclear states.  

For a one-body operator $\hat{O}$,
Eq. (\ref{eq1}) is exact and factors the operator physics -- embodied in the
single-particle matrix elements $\langle |\alpha|;{\textstyle {1 \over 2}} \vdots \vdots  O_{J;T} \vdots \vdots |\beta|; {\textstyle {1 \over 2}} \rangle$ --
from the many-body nuclear physics, which is contained within the one-body density matrix
\begin{equation}
\psi_{J T}^{f,i}(\alpha,\beta) ={1 \over [J][T]} \langle J_f ; T_f \vdots \vdots [c^\dagger_{| \alpha |} \otimes  \tilde{c}_{| \beta |} ]_{J;T} \vdots \vdots J_i ; T_i \rangle.
\label{eqdm}
\end{equation}
Here the single-nucleon creation and annihilation operators carrying good angular momentum and
isospin are defined by
\begin{eqnarray}
c^\dagger_\alpha &=&c^\dagger_{|\alpha|  m_\alpha ; m_{t_\alpha}} \nonumber \\
\tilde{c}_\alpha &=& (-1)^{j_\alpha-m_\alpha + 1/2-m_{t_\alpha}} c_{|\alpha| -m_\alpha; -m_{t_\alpha}},
\end{eqnarray}
where $m_\alpha$ and $m_{t_\alpha}$ are the single-nucleon magnetic quantum numbers for
angular momentum and isospin and $\alpha = \{|\alpha| m_\alpha; m_{t_\alpha} \} $.

In practice, of course, the density matrix of Eq. (\ref{eqdm}) is often not calculated from first principles:
for the classical nuclear physics problem of bound nucleons interacting via a two-body
(or two- plus three-body) potential, the Schr\"{o}rdinger-equation many-body problem may not be
tractable.  Instead, 
wave functions are taken from nuclear models, such as the shell model, where
model assumptions restrict the single-particle states that can be occupied.  An effective
interaction, often determined empirically, is employed to account for the effects of the
omitted degrees of freedom.  This leads to a finite
basis of Slater determinants for the many-body problem and thus a Hamiltonian problem that can
be solved by direct diagonalization.

To take the example we will use in this paper, consider the case of a $0 \hbar \omega$ shell-model
calculation for $^{12}$C.  That is, the included space consists of all Slater determinants where
the $1s$ shell is fully occupied and eight valence nucleons are distributed in all allowed ways
among the twelve single-particle orbitals within the $1p$ shell.  Initial and final nuclear states
determined from such a calculation would both carry positive parity.   Under this assumption the density
matrix for a transition between two such states simplifies,
\begin{widetext}
\begin{eqnarray}
&&\{\psi_{JT}^{f,i}(\alpha,\beta) \} \rightarrow 
\{ \psi_{JT}^{f,i}(1p_{1/2},1p_{1/2}),
\psi_{JT}^{f,i~+}(1p_{3/2},1p_{1/2}), 
\psi_{JT}^{f,i~-}(1p_{3/2},1p_{1/2}),
\psi_{JT}^{f,i}(1p_{3/2},1p_{3/2}),
\psi_{JT}^{f,i}(1s_{1/2},1s_{1/2}) \}; \nonumber \\
&& \psi_{JT}^{f,i~\pm}(1p_{3/2},1p_{1/2}) \equiv \psi_{JT}^{f,i}(1p_{3/2},1p_{1/2}) \pm \psi_{JT}^{f,i}(1p_{1/2},1p_{3/2});~~~ \psi_{JT}^{f,i}(1s_{1/2},1s_{1/2})= 2 \delta_{f,i}
\delta_{J,0}  \delta_{T,0}
\label{dm}
\end{eqnarray}
\end{widetext}
where $\psi_{JT}^{f,i~\pm}(1p_{3/2},1p_{1/2})$ are defined in anticipation of the time-reversal properties of the operators we will discuss below.  Note that the $1s_{1/2}$ shell only contributes to
$J$=0 and $T$=0 elastic transitions. 

So far the single-particle basis has not been specified: in many cases it can remain
unspecified through the calculation of the one-body density matrix.  This would
be the case in shell-model calculations where matrix elements of the effective interaction
are treated as parameters, determined by fitting energy levels, e.g., as Cohen and Kurath \cite{CK}
did in the $1p$ shell.
Such a shell model calculation would yield specific numerical values for the density
matrix elements of Eq. (\ref{dm}).  However, when Eq. (\ref{eq1}) is invoked to evaluate
matrix elements, generally a single-particle basis must be specified.

The harmonic oscillator basis is an attractive choice for this purpose 
because the single-particle matrix elements for electroweak interactions can be evaluate
analytically and have a simple form involving polynomials in the magnitude of the three-momentum transfer
to the nucleus.   This is very useful for determining the functional
form of electroweak response functions.  Even if the density matrix is unknown, response
functions can be expressed in a form where the density matrix elements are parameters.
This allows one to impose experimental constraints on model calculations -- to quickly
determine what degrees of freedom in the density matrix are already constrained by 
experiment, so that these can be eliminated, reducing the model dependence in predicting
other process.  In an example we will discuss later,
charged- and neutral-current neutrino reactions in $^{12}$C involving the triad of 
$J^\pi T = 1^+1$, $M_T = (-1,0,1)$ excited states, the density matrix elements are sharply constrained
by results from electron scattering (and gamma decay), $\beta$ decay, and $\mu$ capture \cite{dubach}.
In some truncated model spaces
there may be a sufficient number of experimental constraints  to determine all of the contributing 
density matrix elements:
this would then free one from dealing with nuclear models.  
 
\section{The basic nuclear operators and their harmonic oscillator matrix elements}
\label{basic}
The semileptonic weak nuclear operators $\hat{O}_{JM;TM_T}$ of interest in this paper
are one-body and can be expressed as a product
of space-spin and isospin operators.  In first quantization
\beq
\sum_{i=1}^A O_{JM}^T(i) I_{TM_T}(i),
\eeq
where the space-spin operator $O_J^T$ includes an isospin label because it contains couplings,
like magnetic moments, that have different strengths depending on the isospin, e.g., the
isoscalar magnetic moments is not equal to the isovector magnetic moment.  Semileptonic 
nuclear cross sections and rates can be expressed in terms of nuclear matrix elements reduced
in angular momentum
\begin{eqnarray}
&&\langle J_f || \hat{O}_{J;TM_T} || J_i \rangle \rightarrow
\langle J_f; T_f M_{T_f} || \hat{O}_{J;TM_T} || J_i; T_i M_{T_i} \rangle = \nonumber \\
&&(-1)^{T_f-M_{T_f}} \left( \begin{array}{ccc}  T_f & T & T_i \\ -M_{T_f} & M_T & M_{T_i} \end{array} \right) 
\sum_{|\alpha|,|\beta|}
\psi_{J T}^{f,i}(|\alpha|,|\beta|) \langle |\alpha| ||  O_J^T || |\beta| \rangle \langle {\textstyle {1 \over 2}} || I_T || {\textstyle {1 \over 2}} \rangle,
\label{eq:isospin}
\end{eqnarray}
where the arrow in the first line indicates we are ignoring small Coulomb and other charge-independence-breaking
terms, so that nuclear states can be labeled as eigenstates of good isospin (a common
assumption in nuclear structure calculations).  The second line follows from Eq. (\ref{eq1}).

The one-body space-spin operators $O^T_J$ governing
semileptonic electroweak interactions are derived in standard references \cite{walecka} by
expanding the vector and axial-vector charge and current operators through order $|\vec{p}|/M$.
They can be expressed in terms of the seven single-particle operators
\begin{eqnarray}
\label{omegaj}
 M^{M_J}_J(\qx) \nonumber \\
\Delta^{M_J}_J(\qx)&\equiv& \bm^{M_J}_{J J}(\qx) \cdot \frac{1}{q} \vec{\bnab} \nonumber \\
\Delta^{\prime M_J}_J(\qx)&\equiv&  -i \left[\frac{1}{q} \vec{\bnab} \times \bm^{M_J}_{J J}(\qx) \right]\cdot \frac{1}{q} \vec{\bnab}
= \left[J \right]^{-1} \left[- J^{1/2} \bm^{M_J}_{J J+1}(\qx)
+ (J+1)^{1/2}\bm^{M_J}_{J J-1} (\qx) \right]\cdot \frac{1}{q} \vec{\bnab} \nonumber \\
\Sigma^{M_J}_J(\qx)&\equiv& \bm^{M_J}_{J J}(\qx) \cdot \vec{\bsig} \nonumber \\
\Sigma^{\prime M_J}_J(\qx)&\equiv&  -i \left[\frac{1}{q} \vec{\bnab} \times \bm^{M_J}_{J J}(\qx) \right]\cdot  \vec{\bsig} 
= \left[J \right]^{-1} \left[- J^{1/2} \bm^{M_J}_{J J+1}(\qx)
+  (J+1)^{1/2}\bm^{M_J}_{J J-1} (\qx) \right]\cdot  \vec{\bsig} \nonumber \\
\Sigma^{\prime\prime M_J}_J(\qx)&\equiv& \left[\frac{1}{q} \vec{\bnab}  M^{M_J}_{J }(\qx) \right]\cdot  \vec{\bsig} 
= \left[J \right]^{-1} \left[(J+1)^{1/2} \bm^{M_J}_{J J+1}(\qx)
+  J^{1/2}\bm^{M_J}_{J J-1} (\qx) \right]\cdot  \vec{\bsig} \nonumber \\
\Omega^{M_J}_J(\qx)&\equiv& M^{M_J}_{J }(\qx) ~ \vec{\bsig}  \cdot \frac{1}{q} \vec{\bnab}, 
\end{eqnarray}
where $q$ is the magnitude of the three-momentum transferred to the nucleus and
$[J] \equiv \sqrt{2J+1}$.   Because of operator time-reversal properties it is helpful
to replace $\Omega$ by a new operator
\begin{equation}
\Omega^{\prime M_J}_J(\qx)\equiv \Omega^{M_J}_J(\qx) + \frac{1}{2}\Sigma^{\prime\prime M_J}_J(\qx).
\label{omegaprime}
\end{equation}
The multipole operators in Eqs. (\ref{omegaj}) are constructed from spherical Bessel functions,
spherical harmonics, and vector spherical harmonics \cite{edmonds} 
\begin{eqnarray}
M^{M_J}_J(\qx) &\equiv& j_J (qx) Y^{M_J}_J(\Omega_x) \nonumber \\
\bm^{M_J}_{JL}(\qx)&\equiv& j_L (qx) \by^{M_J}_{JL1}(\Omega_x)
\label{mjldef}
\end{eqnarray}
where
\begin{equation}
\by^{M_J}_{JL1}(\Omega_x)=[Y_L \otimes \be_1]_{J M_{J}} 
=\sum_{m \lambda} \langle L m 1 \lambda | (L 1)J M_J \rangle Y_L^m(\Omega_x) \be_{1 \lambda}.
\label{harmonicdef}
\end{equation}
The spherical unit vectors $\be_{1 \lambda}$  are defined by $\be_{1 \pm 1} = \mp (\be_x \pm i \be_y)/\sqrt{2}$ and
$\be_{1 0}=\be_z$.  

The single-particle basis $|\alpha \rangle = |n_\alpha (l_\alpha 1/2)j_\alpha m_\alpha
\rangle$ of Eq. (\ref{eq1}) has the general coordinate-space form
\begin{equation}
R_{n_\alpha l_\alpha j_\alpha}(x) [Y_{l_\alpha}(\Omega_x) \otimes \xi_{1/2}]_{j_\alpha m_\alpha}
\end{equation}
where $\xi_{1/2}$ is the Pauli spinor.  The angular and spin portions of the single-particle
operator matrix elements can then be performed, leaving only matrix elements between
unspecified radial wave functions,

\begin{widetext}
\beq
&&\langle n^\prime (l^\prime {\textstyle{1 \over 2}})j^\prime|| M_J(\qx) || n (l {\textstyle{1 \over 2}})j\rangle= {1 \over\sqrt{ 4 \pi}}(-)^{J+j+1/2} \left[l^\prime\right] \left[l\right]  \left[j^\prime \right]  \left[j \right]  \left[J\right] 
\left\{ \begin{array}{ccc}
l^\prime  & j^\prime  & {\textstyle {1 \over 2}}  \\
j & l  & J  \end{array} \right\}
\left( \begin{array}{ccc}
l^\prime  &J  & l \\
0 & 0 & 0  \end{array} \right)
 \langle \np \lp\jp| j_J(\rho)| n l j \rangle 
\label{mjbracket}\nonumber\\ \nonumber
&&\langle n^\prime (l^\prime {\textstyle{1 \over 2}})j^\prime|| \bm_{J L}(\qx)\cdot  \vec{\bsig}  || n (l {\textstyle{1 \over 2}})j\rangle= \sqrt{{3 \over 2 \pi}}(-)^{l^\prime} \left[l^\prime \right] \left[l\right]  \left[j^\prime \right]  \left[j \right]  \left[L\right] \left[J\right] 
\left\{ \begin{array}{ccc}
\lp & l  & L  \\
{\textstyle {1 \over 2}}  & {\textstyle {1 \over 2}}   & 1  \\
\jp  & j  & J  \end{array} \right\}
\left( \begin{array}{ccc}
\lp  & L & l \\
0 & 0 & 0  \end{array} \right)
 \langle \np \lp\jp| j_L(\rho)| n l j \rangle 
\label{mjlsigmabracket}\nonumber\\
&&\langle n^\prime (l^\prime {\textstyle{1 \over 2}})j^\prime|| \bm_{J L}(\qx)\cdot  \frac{1}{q} \vec{\bnab} || n (l {\textstyle{1 \over 2}})j\rangle= {1 \over \sqrt{4 \pi}}(-)^{L+j+1/2} \left[l^\prime \right] \left[j^\prime\right]  \left[j \right]  \left[L \right]  \left[J\right] 
\left\{ \begin{array}{ccc}
\lp  & \jp   & {\textstyle {1 \over 2}}  \\
j  & l  & J  \end{array} \right\}\nonumber\\
&&\hskip 3truecm \times \Biggl\{ -(l+1)^{1/2} \left[l+1 \right]   
\left\{ \begin{array}{ccc}
L   & 1  & J \\
l & \lp & l+1
\end{array} \right\}
\left( \begin{array}{ccc}
\lp & L  & l+1 \\
0  &  0  &  0
 \end{array} \right)
             \langle \np \lp\jp| j_L(\rho) \left( \frac{d}{d\rho} - \frac{l}{\rho}\right) | n l j \rangle \nonumber\\ 
&&\hskip 3.8truecm +  l^{1/2} \left[l-1 \right] 
\left\{ \begin{array}{ccc}
L  &  1 & J \\
l  &  \lp  & l-1
\end{array} \right\}
\left( \begin{array}{ccc}
\lp  & L  & l-1 \\
0 & 0 & 0 \end{array} \right)
             \langle \np \lp\jp| j_L(\rho) \left( \frac{d}{d\rho} + \frac{l+1}{\rho}\right) | n l j \rangle  \Biggr\} 
\label{mjlnablabracket}\nonumber\\
&&\langle n^\prime (l^\prime {\textstyle{1 \over 2}})j^\prime|| M_J(\qx) \vec{\bsig} \cdot  \frac{1}{q} \vec{\bnab} || n (l {\textstyle{1 \over 2}})j\rangle= {1 \over \sqrt{4 \pi}}(-)^{l^\prime} \left[l^\prime\right] \left[j^\prime \right]  \left[j \right]  \left[2j - l \right]  \left[J\right]
\left\{ \begin{array}{ccc}
\lp & \jp  &  {\textstyle {1 \over 2}} \\
j & 2j-l  & J  \end{array} \right\}
\left( \begin{array}{ccc}
\lp &  J&  2j-l \\
0  & 0 & 0  \end{array} \right) \nonumber\\
%
&&\hskip 3truecm  \times \Biggl\{ -\delta_{j,l+1/2}   \langle \np \lp\jp| j_J(\rho) \left( \frac{d}{d\rho} - \frac{l}{\rho}\right) | n l j \rangle   
+  \delta_{j,l- 1/2} \langle \np \lp\jp| j_J(\rho) \left( \frac{d}{d\rho} + \frac{l+1}{\rho}\right) | n l j \rangle      \Biggr\}.
\label{mjnablabracket}
\eeq

\end{widetext}
Here the dimensionless coordinate $\rho \equiv q x$, while the radial matrix elements are defined by
\begin{equation}
\langle n' l' j' | \theta(\rho) | n l j \rangle \equiv \int x^2 dx R^*_{n'l'j'}(x) \theta(\rho) R_{n l j}(x)
\end{equation}
for 
\begin{eqnarray}
\theta(\rho) = \left\{ \begin{array}{l} j_J(\rho) \\ j_J(\rho) \left({d \over d\rho}-{l \over \rho} \right) \\
j_J(\rho) \left({d \over d\rho} + {l+1 \over \rho} \right)~. \end{array} \right.
\label{jmes}
\end{eqnarray}

To complete the evaluation a specific choice for the single-particle radial wave functions must
be made.  The harmonic oscillator is an attractive choice, convenient both for nuclear structure
structure reasons (for example, in certain model spaces spurious center-of-mass degrees of
freedom can be removed exactly) and because the radial matrix elements of Eqs. (\ref{jmes})
can be evaluated analytically.  The properly normalized radial wave functions are
\begin{equation}
R_{nlj}(x) \equiv R_{nl}(x) = \left[ {2 e^z \over b^3 (n-1)! \Gamma(n+l+{\textstyle {1 \over 2}}) z^{l+1}} \right]^{1/2}
 {d^{n-1} \over dz^{n-1}} \left[ z^{n+l-1/2} e^{-z} \right],
\end{equation}
where $z \equiv (x/b)^2$ and $b$ is the oscillator size parameter.  We employ a notation
where the nodal quantum number $n = (N-l)/2+1$, where the principal quantum number $N$ = 0,1,2,...
Thus $n$= 1,2,3...   A given harmonic oscillator state can be labeled by $n,l,j$ or,
equivalently, $N,j$:  as $N$ determines the parity of the shell, knowledge of $N$ and $j$ uniquely
determines $l$.  Our Mathematica script uses $N,j$ to identify single-nucleon states.   

The radial integrals appearing in Eqs. (\ref{mjnablabracket}) can be evaluated analytically
for harmonic oscillator states,
\begin{widetext}
\begin{eqnarray}
&&\langle n' l' | j_L(\rho) | n l \rangle = {2^L \over (2L+1)!!} y^{L/2} e^{-y}
\sqrt{(n-1)! (n'-1)! \Gamma(n'+l'+{\textstyle {1 \over 2}}) \Gamma(n+l+{\textstyle {1 \over 2}})}
\sum_{k=0}^{n-1} \sum_{k'=0}^{n'-1}
{(-1)^{k+k'} \over k! k'!} \nonumber \\
&&~~~~ \times~ {1 \over (n-1-k)! (n'-1-k')!}
{\Gamma[{\textstyle {1 \over 2}} (l+l'+L+2k+2k'+3)] \over
\Gamma[l+k+{\textstyle {3 \over 2}}] \Gamma[l'+k'+{\textstyle {3 \over 2}}]} ~{}_1 F_1[{\textstyle {1 \over 2}}(L-l-l'-2k-2k');L+{\textstyle{3 \over 2}};y] \\
&&\langle n' l' | j_L(\rho) ({d \over dr} - {l \over r}) | n l \rangle ={2^{L-1} \over (2L+1)!!} y^{(L-1)/2} e^{-y}
\sqrt{(n-1)! (n'-1)! \Gamma(n'+l'+{\textstyle {1 \over 2}}) \Gamma(n+l+{\textstyle {1 \over 2}})} \nonumber \\
&&~~~~\times ~\sum_{k=0}^{n-1} \sum_{k'=0}^{n'-1}
{(-1)^{k+k'} \over k! k'!} {1 \over (n-1-k)! (n'-1-k')!} {\Gamma[{\textstyle {1 \over 2}}(L+l+l'+2k+2k'+2)] \over
\Gamma[l+k+{\textstyle {3 \over 2}}] \Gamma[l'+k'+{\textstyle {3 \over 2}}]} 
\Biggl\{ -{\textstyle{1 \over 2}}(l+l'+L+2k+2k'+2)  \nonumber \\ 
&& ~~~~ \times ~{}_1 F_1[{\textstyle {1 \over 2}} (L-l-l'-2k-2k'-1);
L+{\textstyle {3 \over 2}};y] + (2k) {}~_1 F_1[{\textstyle {1 \over 2}}(L-l-l'-2k-2k'+1);L+{\textstyle {3 \over 2}};y] \Biggr\} \\
&&\langle n' l' | j_L(\rho) ({d \over dr} + {l+1 \over r}) | n l \rangle ={2^{L-1} \over (2L+1)!!} y^{(L-1)/2} e^{-y}
\sqrt{(n-1)! (n'-1)! \Gamma(n'+l'+{\textstyle {1 \over 2}}) \Gamma(n+l+{\textstyle {1 \over 2}})} \nonumber \\
&&~~~~\times~ \sum_{k=0}^{n-1} \sum_{k'=0}^{n'-1}
{(-1)^{k+k'} \over k! k'!} {1 \over (n-1-k)! (n'-1-k')!} {\Gamma[{\textstyle {1 \over 2}}(L+l+l'+2k+2k'+2)] \over
\Gamma[l+k+{\textstyle {3 \over 2}}] \Gamma[l'+k'+{\textstyle {3 \over 2}}]} 
\Biggl\{ -{\textstyle {1 \over 2}}(l+l'+L+2k+2k'+2) \nonumber \\ 
&& ~~~~ \times~{}_1 F_1[{\textstyle {1 \over 2}} (L-l-l'-2k-2k'-1);
L+{\textstyle {3 \over 2}};y] + (2l+2k+1) {}~_1 F_1[{\textstyle {1 \over 2}} (L-l-l'-2k-2k'+1);L+{\textstyle {3 \over 2}};y] \Biggr\}
\end{eqnarray}
\end{widetext}
where $ y = (q b/2)^2$. Here  ${}_1F_1$ is the confluent hypergeometric function
\begin{equation}
{}_1F_1(\alpha; \beta; y) = 1 + {\alpha \over \beta} + {\alpha(\alpha+1) \over \beta (\beta+1)}
{y^2 \over 2!} + ...
\end{equation}
In the present application $\alpha$ is a nonpositive integer, so that this series is a polynomial
of order $-\alpha$ in $y$.

Thus the single-nucleon harmonic-oscillator matrix elements of the seven basic operators are completely determined in terms of simple functions.  One finds
\begin{equation}
\langle n^\prime (l^\prime {\textstyle {1 \over 2}})j^\prime|| T_{J}(\qx)  || n (l {\textstyle {1 \over 2}})j\rangle = 
{1 \over \sqrt{4 \pi}}  y^{(J-K)/2} e^{-y} p(y)~,
\label{analytical}
\end{equation}
where $K=2$ for the normal parity operators $M,\Delta^\prime, \Sigma$; $K=1$ for the abnormal parity operators $\Delta, \Sigma^{\prime},\Sigma^{\prime\prime}, \Omega$ (or $\Omega^\prime$).
The tabulations of Ref. \cite{donnelly} provide the polynomials $p(y)$ for such matrix elements,
with the restrictions described previously.  The Mathematica script 
accompanying this paper generates the full expression in Eq. (\ref{analytical}), without restrictions
on the quantum numbers and in a form that can be easily manipulated within Mathematica
to generate analytical expressions for form factors, rates, etc.

An important property related to the time-reversal properties of these operators is
\begin{equation}
\langle n^\prime (l^\prime {\textstyle {1 \over 2}})j^\prime|| T_{J}(\qx)  || n (l {\textstyle {1 \over 2}})j\rangle = 
(-1)^\lambda \langle n (l {\textstyle {1 \over 2}})j|| T_{J}(\qx)  || n^\prime (l^\prime {\textstyle {1 \over 2}})j^\prime \rangle
\end{equation}
where $\lambda=j^\prime-j$ for the operators $M$, $\Delta$, $\Sigma^\prime$, and
$\Sigma^{\prime \prime}$, and $\lambda=j^\prime+j$ for the operators $\Delta^\prime$,
$\Sigma$, and $\Omega^\prime$.  The operator $\Omega^\prime$ of Eq. (\ref{omegaprime})
was defined because $\Omega$ lacks such a simple transformation property.  This
``turn around'' property is important because it leads to a dependence of observables
of density matrix element combinations such as $\psi_{J,T}^{f,i~\pm}(1p_{3/2},1p_{1/2})$.\\

\section{Relationship with semileptonic processes in nuclei}
\label{relationship}

To illustrate the use of the Mathematica script, we will consider various charged-current and neutral-current neutrino 
reactions from the $J^\pi T=0^+0$ ground state of $^{12}$C:
\begin{eqnarray}
{}^{12}C(g.s.)(\nu_e,e^-)^{12}N(g.s.) &~~~~~~~&
{}^{12}C(g.s.)(\bar{\nu}_e,e^+)^{12}B(g.s.) \nonumber \\
{}^{12}C(g.s.)(\nu,\nu)^{12}C(15.11~\mathrm{MeV}) &~~~~~~~&
{}^{12}C(g.s.)(\bar{\nu},\bar{\nu})^{12}C(15.11~\mathrm{MeV}) \nonumber \\
{}^{12}C(g.s.)(\nu,\nu)^{12}C(12.7~\mathrm{MeV}) &~~~~~~~&
{}^{12}C(g.s.)(\bar{\nu},\bar{\nu})^{12}C(12.7~\mathrm{MeV}) \nonumber
\end{eqnarray}
The first four transitions are isovector, populating the triad of $J^+T=1^+1$ states formed by the ground states of $^{12}$N ($M_T$ = 1) and $^{12}$B ($M_T$ = -1) and the 15.11 MeV ($M_T$ = 0) isovector analog state in $^{12}$C.  The last two are isoscalar, populating the $J^+T=1^+0$ 12.7 MeV state
in $^{12}$C.  The four-momentum transfer for these reactions is defined by
\begin{equation}
q^\mu = P_i^\mu-P_f^\mu = k^\mu-k_\nu^\mu = (\omega, \vec{q})
\end{equation}
where $P_i^\mu$, $P_f^\mu$, $k_\nu^\mu$, and $k^\mu$ are the four-momenta of the initial nucleus, final nucleus, incident neutrino, and scattered lepton (electron/positron or neutrino/antineutrino).  The charged-current cross section in the extreme relativistic limit can be written \cite{walecka}
\begin{widetext}
\begin{eqnarray}
\left( {d\sigma \over d\Omega} \right)_{\stackrel{\scriptstyle{\nu_e}}{\scriptstyle{\bar{\nu}_e}}} &=& {2 \over \pi} G_F^2 \cos^2{\theta_c}  F(Z,\epsilon) {\epsilon^2 \over 2J_i+1} \cos^2{{\theta \over 2}} \Bigg\{ \sum_{J=0}^\infty |\langle J_f || \hat{\mathcal{M}}_J + {\omega \over q} \hat{\mathcal{L}}_J || J_i \rangle|^2  \nonumber \\
&+& \left[-{q_\mu^2 \over 2 q^2} + \tan^2{{\theta \over 2}} \right] \sum_{J=1}^\infty \left[ | \langle J_f || \hat{\mathcal{T}}^{el}_J || J_i \rangle|^2 + |\langle J_f || \hat{\mathcal{T}}^{mag}_J || J_i \rangle |^2 \right]  \nonumber \\
&\mp& 2 \tan{{\theta \over 2}} \left[- {q_\mu^2 \over q^2} + \tan^2{{\theta \over 2}} \right]^{1/2} \sum_{J=1}^\infty Re \left(\langle J_f ||\hat{\mathcal{T}}^{mag}_J || J_i \rangle \langle J_f || \hat{\mathcal{T}}_J^{el} || J_i\rangle^* \right) \Biggr\}
\label{eq:cc}
\end{eqnarray}
\end{widetext} 
where $G_F \cos{\theta_c}$ is the weak coupling constant, $\theta$ the angle between the incident neutrino and outgoing electron, $q_\mu^2 = \omega^2-q^2$, and $\epsilon$ is the energy of the outgoing electron/positron.  The function $F(Z,\epsilon)$ corrects the electron phase space for the effects of Coulomb distortion in the field of the daughter nucleus of charge $Z.$  This expression assumes $\epsilon >> m_e$ and neglects nuclear recoil.

The operators $\hat{\mathcal{M}}_J$, $\hat{\mathcal{L}}_J$, $\hat{\mathcal{T}}^{el}$, and $\hat{\mathcal{T}}^{mag}$ are the familiar charge, longitudinal, transverse electric, and transverse magnetic projections of the weak charge-changing hadronic current,
\begin{equation}
\hat{\mathcal{J}}_\mu^\pm = \hat{\mathcal{J}}^1_\mu \pm \hat{\mathcal{J}}^2_\mu
\label{eq:pm}
\end{equation}
where $\hat{\mathcal{J}}_\mu^i$, $i=1,2,3$ are the three components of an isovector.  The weak
current is made up of vector and axial-vector components
\begin{equation}
\hat{\mathcal{J}}_\mu^i = \hat{J}_\mu^i + \hat{J}_\mu^{5,i}.
\label{eq:weakcurrent}
\end{equation}

Multipole projections of this current are taken to exploit the angular momentum and parity quantum labels that are normally carried by the initial and final nuclear states.  
The multipole operators are defined by
\begin{eqnarray}
\hat{\mathcal{M}}_{J M_J; T M_T} &\equiv& \int d {\bf x}~{\bf M}_J^{M_J} (q {\bf x}) {\hat {\mathcal{J}}}_0({\bf x})_{T M_T}
=  \hat{M}_{J M_J; T M_T} + \hat{M}_{J M_J; T M_T}^5 \nonumber \\
\hat{\mathcal{L}}_{J M_J; T M_T} &\equiv& {i \over q}  \int d {\bf x}  \left[ \vec{{\bf \nabla}} M_J^{M_J}(q {\bf x}) \right] \cdot {\hat {\bf {\mathcal{J}}}}({\bf x})_{T M_T} 
= \hat{L}_{J M_J; T M_T} + \hat{L}_{J M_J; T M_T}^5 \nonumber \\
\hat{\mathcal{T}}^{el}_{J M_J; T M_T} &\equiv& {1 \over q}  \int d {\bf x}  \left[ \vec{{\bf \nabla}} \times \bm_{JJ}^{M_J}(q {\bf x}) \right] \cdot {\hat {\bf {\mathcal{J}}}}({\bf x})_{T M_T}
= \hat{T}^{el}_{J M_J; T M_T} + \hat{T}^{el5}_{J M_J; T M_T} \nonumber \\
\hat{\mathcal{T}}^{mag}_{J M_J; T M_T} &\equiv&  \int d {\bf x}  \bm_{JJ}^{M_J}(q {\bf x}) \cdot {\hat {\bf {\mathcal{J}}}}({\bf x})_{T M_T} 
= \hat{T}^{mag}_{J M_J; T M_T} + \hat{T}^{mag5}_{J M_J; T M_T}
\label{eq:mults4} 
\end{eqnarray}
where we have separated the vector and axial-vector (with superscript `5') contributions in Eqs. (\ref{eq:mults4}).  The operators $\hat{M}_J$, $\hat{L}_J$, $\hat{T}_J^{el}$, and
$\hat{T}_J^{mag5}$ are normal parity operators, that is, $\Delta \pi = (-1)^J$.  The
operators $\hat{M}_J^5$, $\hat{L}_J^5$, $\hat{T}_J^{el5}$, and $\hat{T}_J^{mag}$ are 
abnormal parity, $\Delta \pi = (-1)^{J+1}$.   Also note that, for a conserved vector current,
$\hat{L}_J$ can be eliminated as
\begin{equation}
\hat{L}_{JM_J;TM_T} = {\omega \over q} \hat{M}_{JM_J;TM_T}.
\end{equation}\\

Although contributions from two-body 
(exchange) currents can be included, more commonly the nuclear charges and currents
are modeled as the sum of the one-body contributions from the constituent protons and
neutrons.   For a single free nucleon one has the following general forms for the matrix 
elements of the vector and axial-vector isovector currents for charge-changing weak interactions
\begin{widetext}
\begin{eqnarray}
\langle \vec{k}' \lambda'; {\textstyle {1 \over 2}} m_{t'} | J_\mu^\pm (0) | \vec{k} \lambda; {\textstyle {1 \over 2}} m_t \rangle &=&
\bar{u}(\vec{k}' \lambda') \{F_1^{(1)}(q_\mu^2) \gamma_\mu -i F_2^{(1)}(q_\mu^2) \sigma_{\mu \nu} q^{\nu}
\} u(\vec{k} \lambda) \langle {\textstyle {1 \over 2}} m_{t'} | \tau_\pm | {\textstyle {1 \over 2}} m_t \rangle \nonumber \\
\langle \vec{k}' \lambda'; {\textstyle {1 \over 2}} m_{t'} | J_\mu^{5\pm} (0) | \vec{k} \lambda; {\textstyle {1 \over 2}} m_t \rangle &=&
\bar{u}(\vec{k}' \lambda') \{F_A^{(1)}(q_\mu^2) \gamma_\mu \gamma_5 - F_P^{(1)}(q_\mu^2) \gamma_5 q_\mu \} u(\vec{k} \lambda) \langle {\textstyle {1 \over 2}} m_{t'} | \tau_\pm | {\textstyle {1 \over 2}} m_t \rangle
\label{eq:currentsA}
\end{eqnarray}
\end{widetext} 
where we have omitted second-class scalar (which violates conservation of the vector
current) and axial-tensor couplings.  [Our notation in this paper follows Bjorken and Drell
\cite{BD}, including gamma matrices, metric, and current definitions, with the exception of the definition of the form factors, which follows that of Ref. \cite{walecka}.]

A standard 
non relativistic reduction of the matrix elements through order $1/M_N$, in which
momenta are interpreted as gradients operating within the nucleus, leads to expressions
for the multipole operators of Eqs. (\ref{eq:mults4}) in terms of the seven basic single-particle operators defined in
Eqs. (\ref{omegaj}):
\begin{eqnarray}
M_{JM_J}^\pm(\qx) &=& F_1^{(1)}(q_\mu^2) M_J^{M_J}(\qx) \tau_\pm \nonumber \\
T^{el~\pm}_{JM_J}(\qx) &=& 
{q \over M_N} (F_1^{(1)}(q_\mu^2) \Delta^{'M_J}_J (\qx) + {\textstyle {1 \over 2}} \mu^{(1)}(q_\mu^2) \Sigma^{M_J}_J(\qx)) \tau_\pm  \nonumber \\
T^{mag~\pm}_{JM_J}(\qx) &=&
-{iq \over M_N} (F_1^{(1)}(q_\mu^2) \Delta_J^{M_J}(\qx) - {\textstyle {1 \over 2}} \mu^{(1)}(q_\mu^2) \Sigma_J^{'M_J}(\qx)) \tau_\pm  \nonumber \\
M^{5~\pm}_{JM_J}(\qx) &=&
{iq \over M_N} (F_A^{(1)}(q_\mu^2) \Omega_J^{'M_J}(\qx) + {\textstyle {1 \over 2}} \omega F_P^{(1)}(q_\mu^2) \Sigma^{''M_J}_J(\qx)) \tau_\pm \nonumber \\
L_{JM_J}^{5~\pm}(\qx) &=& 
i\left(F_A^{(1)}(q_\mu^2)-{q^2 \over 2M_N} F_P^{(1)}(q_\mu^2) \right) \Sigma^{''M_J}_J(\qx) \tau_\pm \nonumber \\
T_{JM_J}^{el5~\pm}(\qx) &=& iF_A^{(1)}(q_\mu^2) \Sigma^{'M_J}_J(\qx) \tau_\pm \nonumber \\
T_{JM_J}^{mag5~\pm}(\qx) &=& F_A^{(1)}(q_\mu^2) \Sigma^{M_J}_J(\qx) \tau_\pm 
\label{eq:final}
\end{eqnarray}
where $M_N$ is the nucleon mass and $\mu^{(1)}(q_\mu^2) = F_1^{(1)}(q_\mu^2) + 2M_N F_2^{(1)}(q_\mu^2)$.

The nucleon isovector form factors appearing in Eq. (\ref{eq:final}) include
the charge form factor ($F_1^{(1)}(0)=1$),
the magnetic moment ($\mu^{(1)}(0) \sim 4.706$), and the axial form factor ($F_A^{(1)}(0) \sim -1.26$).  Various dipole and other representations of the momentum dependence of $F_1^{(1)}$,
$F_2^{(1)}$, and $F_A^{(1)}$ are available; the momentum dependence of the pseudoscalar form factor $F_P^{(1)}$ is found to be
\begin{equation}
F_P^{(1)}(q_\mu^2) \sim {2 M_N F_A^{(1)}(0) \over m_\pi^2-q^2},
\end{equation}
assuming pion-pole dominance and the Goldberger-Treiman relationship \cite{walecka}.
The isospin raising/lowering operator $\tau_\pm$ is related to 
the spherical projections of isospin Pauli operator $\tau_{1 m}$ by
\begin{equation}
I_{T=1}^\pm = {\tau_1 \pm \tau_2 \over 2} \equiv \tau_\pm = \mp {1 \over \sqrt{2}} \tau_{1 \pm 1}.
\label{eq:tau}
\end{equation}\\

\noindent
{\it Charged-current neutrino scattering off $^{12}$C:}  Now we consider the example of charged-current neutrino scattering off the $J^\pi T = 0^+0$ 
$^{12}$C ground state, leading to the $1^+1$ grounds states of $^{12}$B and $^{12}$N.
We treat these states as isospin eigenstates, ignoring isospin breaking due to 
electromagnetic or other charge-dependent
interactions. Thus Eq. (\ref{eq:cc}) simplifies because of the
angular momentum and parity constraints imposed by the  $0^+0 \rightarrow 1^+1$ 
transition and because
the matrix elements can be reduced in isospin,
\begin{widetext}
\begin{eqnarray}
\left( {d\sigma \over d\Omega} \right)_{\stackrel{\scriptstyle{^{12}C(\nu_e,e^-)^{12}N(g.s.)}}{\scriptstyle{^{12}C(\bar{\nu}_e},e^+)^{12}B(g.s.)}} &=& {2 \epsilon^2 \over 3 \pi} G_F^2 \cos^2{\theta_c}  F(Z,\epsilon) \cos^2{{\theta \over 2}} \Biggl\{ |\langle 1^+;1\vdots \vdots \hat{M}_{1;1}^5 + {\omega \over q} \hat{L}_{1;1}^5 \vdots \vdots 0^+;0 \rangle|^2 \nonumber \\
&+& \left[-{q_\mu^2 \over 2 q^2} + \tan^2{{\theta \over 2}} \right]  \left[ | \langle 1^+;1 \vdots \vdots \hat{T}^{el5}_{1;1} \vdots \vdots 0^+;0 \rangle|^2 + |\langle 1^+;1 \vdots \vdots \hat{T}^{mag}_{1;1} \vdots \vdots 0^+;0 \rangle |^2 \right]  \nonumber \\
&\mp& 2  \tan{{\theta \over 2}} \left[- {q_\mu^2 \over q^2} + \tan^2{{\theta \over 2}} \right]^{1/2}Re \left(\langle 1^+;1 \vdots \vdots \hat{T}^{mag}_{1;1} \vdots \vdots 0^+;0 \rangle \langle 1+;1 \vdots \vdots \hat{T}_{1;1}^{el5} \vdots \vdots 0^;0 \rangle^* \right) \Biggr\}
\label{eq:cc12}
\end{eqnarray}
\end{widetext} 
where a factor of 1/3 results from isospin reduction,
\begin{equation}
\langle 1; 1 \pm 1  || \hat{O}_{1;1 \pm 1} || 0; 0 0 \rangle = 
{1 \over \sqrt{3}} \langle  1;1   \vdots \vdots \hat{O}_{1;1} \vdots \vdots 0;0  \rangle.
\label{eq1B}
\end{equation}

Eqs. (\ref{eq:cc12}) and (\ref{eq:final}) thus determine the cross section, once the doubly-reduced
single-particle matrix elements are evaluated.   Eqs. (\ref{eq1}) and (\ref{eq:tau})
yield
\begin{eqnarray}
&&\langle 1^+ ; 1 \vdots \vdots \hat{O}_{1;1} \vdots \vdots 0^+; 0 \rangle = 
\sum_{|\alpha|,|\beta|}
\psi_{J=1;T=1}^{f,i}(|\alpha|,|\beta|) \langle |\alpha|;{\textstyle {1 \over 2}} \vdots \vdots  O^{(1)}_{1;1} \vdots \vdots |\beta|;{\textstyle {1 \over 2}} \rangle = \nonumber \\
&&\sum_{|\alpha|,|\beta|}
\psi_{1;1}^{f,i}(|\alpha|,|\beta|) \langle |\alpha|~ ||  O^{(1)}_{J} ||~ |\beta| \rangle ( {\textstyle{ \mp{1 \over \sqrt{2}}}}) \langle {\textstyle {1 \over 2}} || \tau || {\textstyle {1 \over 2}} \rangle =
\mp \sqrt{3} \sum_{|\alpha|,|\beta|}
\psi_{1;1}^{f,i}(|\alpha|,|\beta|) \langle |\alpha|~ ||  O^{(1)}_{J} ||~ |\beta| \rangle
\label{eq1A}
\end{eqnarray}
where we have used $\langle {\textstyle {1 \over 2}} || \tau || {\textstyle {1 \over 2}} \rangle = \sqrt{6}$.  The remaining spin-spatial matrix
element can then be evaluated with the Mathematica program, by virtue of Eq. (\ref{eq:final}).  
As discussed early, we truncate the
density matrix to the $1p$ shell and assume a harmonic oscillator basis, but otherwise keep 
the elements of the density matrix arbitrary: In the next section we will describe how the
Mathematica code can be used to provide the needed matrix element, once this truncation
has been made.  The result is an analytic expression for the cross
section that encompasses any shell-model calculation restricted in this way,
\begin{widetext}
\begin{eqnarray}
&&\left( {d\sigma \over d\Omega} \right)_{\stackrel{\scriptstyle{^{12}C(\nu_e,e^-)^{12}N(g.s.)}}{\scriptstyle{^{12}C(\bar{\nu}_e},e^+)^{12}B(g.s.)}} = {\epsilon^2 \over 2 \pi^2} G_F^2 \cos^2{\theta_c}  F(Z,\epsilon) \cos^2{{\theta \over 2}} e^{-2y} \nonumber \\
&& \Biggl\{
\left[{q \over M} F_A^{(1)} \psi_{31}^+ +{\omega \over q} \left(F_A^{(1)}-{q^2 \over M} F_P^{(1)} \right) \left(-{\sqrt{2} \over 3} (1+2y) \psi_{11} - {4 \over 3} (1-y) \psi_{31}^- + {2 \sqrt{5} \over 3} (1 - {2 \over 5}y)\psi_{33} \right) \right]^2 +
\nonumber \\
&& \left[-{q_\mu^2 \over 2 q^2} + \tan^2{{\theta \over 2}} \right] \Biggl\{ \left[ F_A^{(1)} \left(-{2 \over 3}(1-2y)\psi_{11}
-{4 \sqrt{2} \over 3} (1-{y \over 2}) \psi_{31}^- + {2 \sqrt{10} \over 3} (1-{4 \over 5}y) \psi_{33} \right) \right]^2
+ \nonumber \\
&& {q^2 \over M^2} \left[F_1^{(1)} \left({2 \over 3} \psi_{11} + {\sqrt{2} \over 3} \psi_{31}^- + {\sqrt{10} \over 3} \psi_{33} \right) + {\mu^{(1)} \over 2} \left(-{2 \over 3}(1-2y) \psi_{11} -{4 \sqrt{2} \over 3} (1 -{y \over 2}) \psi_{31}^- + {2 \sqrt{10} \over 3} (1-{4 \over 5}y) \psi_{33} \right) \right]^2 \Biggr\}  
  \nonumber \\
&&\mp 2 \tan{{\theta \over 2}} \left[- {q_\mu^2 \over q^2} + \tan^2{{\theta \over 2}} \right]^{1/2}
F_A^{(1)} {q \over M} \left(-{2 \over 3} (1-2y) \psi_{11} - {4 \sqrt{2} \over 3} (1 - {y \over 2}) \psi_{31}^- + {2 \sqrt{10} \over 3} (1 - {4 \over 5} y) \psi_{33} \right)  \times \nonumber \\
&& \left[F_1^{(1)} \left({2 \over 3} \psi_{11} + {\sqrt{2} \over 3} \psi_{31}^- + {\sqrt{10} \over 3} \psi_{33} \right) + {\mu^{(1)} \over 2} \left( -{2 \over 3}(1-2y)\psi_{11} -{4 \sqrt{2} \over 3} (1-{y \over 2}) \psi_{31}^- + {2 \sqrt{10} \over 3} (1 - {4 \over 5}y) \psi_{33} \right) \right]  \Biggl\}
\label{eq:cc12y}
\end{eqnarray}
\end{widetext} 
where it is understood that all single-nucleon couplings are described by form factors, e.g.,
$F_A^{(1)}=F_A^{(1)}(q_\mu^2)$.  In this expression we used the short-hand notation for the
density matrix of $\psi_{2j_\alpha~2j_\beta} \equiv \psi_{J=1,T=1}^{f=1^+1,i=0^+0}(|\alpha|,|\beta|)$, etc. 
That is, the final and initial nuclear state labels $f,i$ are suppressed, as are the $J,T$ of the
transition, since these quantum numbers are uniquely determined for the transition ($J=1,
T=1$). \\

\noindent
{\it Neutral current scattering to the $1^+1$ 15.11 MeV state of  $^{12}$C:}  
The hadronic neutral current is
\begin{equation}
\hat{\mathcal{J}}_\mu^{NC} = \hat{\mathcal{J}}^3_\mu -2 \sin^2{\theta_W} J_\mu^{em}
\label{eq:neutralcurrent}
\end{equation}
where $\hat{\mathcal{J}}^3_\mu$ is third isospin component of the weak current of Eq. (\ref{eq:pm}), $\hat{\mathcal{J}}_\mu^{em}$ is the electromagnetic current, and $\theta_W$ the Weinberg angle.  Because of the presence of $\hat{\mathcal{J}}_\mu^{em}$, the neutral current includes an isoscalar contribution.  Eqs. (\ref{eq:pm}) and (\ref{eq:neutralcurrent}) allow one to compare the isospin
dependence of charged and neutral-current isovector transitions,
\begin{eqnarray}
I_{T=1}^\pm =  \tau_\pm &\leftrightarrow&  I_{T=1}^{NC} =[1 - 2 \sin^2{\theta_w}] {\tau_3 \over 2},~~{\mathrm{vector~current}}  \nonumber \\
&\leftrightarrow& I_{T=1}^{NC~5} = {\tau_3 \over 2},~~{\mathrm{axial-vector~current}}
\end{eqnarray}
The reduction in isospin between nuclear states (compare Eq. (\ref{eq1B})) yields
\begin{equation}
\langle 1; 1 0  || \hat{O}_{1;10} || 0; 0 0 \rangle = 
{1 \over \sqrt{3}} \langle  1;1   \vdots \vdots \hat{O}_{1;1} \vdots \vdots 0;0  \rangle,
\end{equation}
while the single-particle matrix element for $\tau_3/2$ (compare Eq. (\ref{eq1A})) yields
\begin{eqnarray}
&&\langle 1^+ ; 1 \vdots \vdots \hat{O}^{(1)}_{1;1} \vdots \vdots 0^+; 0 \rangle =
\sum_{|\alpha|,|\beta|}
\psi_{J=1;T=1}^{f,i}(|\alpha|,|\beta|) \langle |\alpha|;{\textstyle {1 \over 2}} \vdots \vdots  O^{(1)}_{1;1} \vdots \vdots |\beta|;{\textstyle {1 \over 2}} \rangle = \nonumber \\
&& {1 \over 2} \sum_{|\alpha|,|\beta|}
\psi_{1;1}^{f,i}(|\alpha|,|\beta|) \langle |\alpha|~ ||  O^{(1)}_{J} ||~ |\beta| \rangle \langle {\textstyle {1 \over 2}} || \tau || {\textstyle {1 \over 2}} \rangle = 
{\sqrt{6} \over 2} \sum_{|\alpha|,|\beta|}
\psi_{1;1}^{f,i}(|\alpha|,|\beta|) \langle |\alpha|~ ||  O^{(1)}_{J} ||~ |\beta| \rangle~.
\end{eqnarray}
Thus there is an overall factor of 1/2 in the cross section due to isospin, relative to charged-current
result in $^{12}$C.  

Finally, the cross section for  neutral-current scattering of neutrinos off nuclei can be obtained from 
Eq. (\ref{eq:cc})
by replacing $\epsilon$ with $\epsilon_\nu$, the energy of the scattered neutrino, and by
removing $F(Z,\epsilon)$ and $\cos^2{\theta_c}$.  It follows 
\begin{widetext}
\begin{eqnarray}
&&\left( {d\sigma \over d\Omega} \right)_{\stackrel{\scriptstyle{^{12}C(\nu,\nu')^{12}C(1^+1)}}{\scriptstyle{^{12}C(\bar{\nu},\bar{\nu}')^{12}C(1^+1)}}} = {\epsilon_\nu^2 \over 4 \pi^2} G_F^2 \cos^2{{\theta \over 2}} e^{-2y} \nonumber \\
&& \Biggl\{
\left[{q \over M} F_A^{(1)} \psi_{31}^+ +{\omega \over q} \left(F_A^{(1)}-{q^2 \over M} F_P^{(1)} \right) \left(-{\sqrt{2} \over 3} (1+2y) \psi_{11} - {4 \over 3} (1-y) \psi_{31}^- + {2 \sqrt{5} \over 3} (1 - {2 \over 5}y)\psi_{33} \right) \right]^2 + 
\nonumber \\
&& \left[-{q_\mu^2 \over 2 q^2} + \tan^2{{\theta \over 2}} \right] \Biggl\{ \left[ F_A^{(1)} \left(-{2 \over 3}(1-2y)\psi_{11}
-{4 \sqrt{2} \over 3} (1-{y \over 2}) \psi_{31}^- + {2 \sqrt{10} \over 3} (1-{4 \over 5}y) \psi_{33} \right) \right]^2
+ {q^2 \over M^2} (1 - 2\sin^2{\theta_W})^2 \times \nonumber \\
&& \left[F_1^{(1)} \left({2 \over 3} \psi_{11} + {\sqrt{2} \over 3} \psi_{31}^- + {\sqrt{10} \over 3} \psi_{33} \right) + {\mu^{(1)} \over 2} \left(-{2 \over 3}(1-2y) \psi_{11} -{4 \sqrt{2} \over 3} (1 -{y \over 2}) \psi_{31}^- + {2 \sqrt{10} \over 3} (1-{4 \over 5}y) \psi_{33} \right) \right]^2 \Biggr\}  
  \nonumber \\
&&\mp 2 \tan{{\theta \over 2}} \left[- {q_\mu^2 \over q^2} + \tan^2{{\theta \over 2}} \right]^{1/2}
F_A^{(1)} {q \over M} (1-2 \sin^2{\theta_W}) \left(-{2 \over 3} (1-2y) \psi_{11} - {4 \sqrt{2} \over 3} (1 - {y \over 2}) \psi_{31}^- + {2 \sqrt{10} \over 3} (1 - {4 \over 5} y) \psi_{33} \right)  \times \nonumber \\
&&  \left[F_1^{(1)} \left({2 \over 3} \psi_{11} + {\sqrt{2} \over 3} \psi_{31}^- + {\sqrt{10} \over 3} \psi_{33} \right) + {\mu^{(1)} \over 2} \left( -{2 \over 3}(1-2y)\psi_{11} -{4 \sqrt{2} \over 3} (1-{y \over 2}) \psi_{31}^- + {2 \sqrt{10} \over 3} (1 - {4 \over 5}y) \psi_{33} \right) \right] \Biggr\}
\label{eq:cc12yNC}
\end{eqnarray}
\end{widetext} 
where, of course, both the Coulomb correction and effects of the Cabibbo angle have
been removed. \\

\noindent
{\it Neutral current scattering to the $1^+0$ 12.7 MeV state  $^{12}$C:}  The transition to the
$1^+0$ 12.7 MeV state is generated by the isoscalar neutral current which, according
to Eq. (\ref{eq:neutralcurrent}), involves
\begin{eqnarray}
&&I_{T=0}^{NC} =-(2 \sin^2{\theta_w}) {1 \over 2},~~{\mathrm{vector~current}}  \nonumber \\
&& I_{T=0}^{NC~5} = 0,~~{\mathrm{axial-vector~current}}
\end{eqnarray}
That is, the isoscalar neutral current is purely vector which, for example in $^{12}$C,
thus limiting the contributing multipoles to $\hat{T}^{mag}_{J=1}$.
The isospin factors are trivial
\begin{equation}
\langle 1; 0 0  || \hat{O}_{1;00} || 0; 0 0 \rangle = 
 \langle  1;0   \vdots \vdots \hat{O}_{1;0} \vdots \vdots 0;0  \rangle,
\end{equation}
while the single-particle matrix element for the unit isospin operator is
\begin{eqnarray}
&&\langle 1^+ ; 0 \vdots \vdots \hat{O}^{(1)}_{1;0} \vdots \vdots 0^+; 0 \rangle =
\sum_{|\alpha|,|\beta|}
\psi_{J=1;T=1}^{f,i}(|\alpha|,|\beta|) \langle |\alpha|; {\textstyle {1 \over 2}} \vdots \vdots  O^{(1)}_{1;1} \vdots \vdots |\beta|;{\textstyle {1 \over 2}} \rangle = \nonumber \\
&&\sum_{|\alpha|,|\beta|}
\psi_{1;1}^{f,i}(|\alpha|,|\beta|) \langle |\alpha|~ ||  O^{(1)}_{J} ||~ |\beta| \rangle \langle {\textstyle {1 \over 2}} || 1 || {\textstyle {1 \over 2}} \rangle =
\sqrt{2} \sum_{|\alpha|,|\beta|}
\psi_{1;1}^{f,i}(|\alpha|,|\beta|) \langle |\alpha|~ ||  O^{(1)}_{J} ||~ |\beta| \rangle
\end{eqnarray}
It follows that the cross section is
\begin{widetext}
\begin{eqnarray}
&&\left( {d\sigma \over d\Omega} \right)_{\stackrel{\scriptstyle{^{12}C(\nu,\nu')^{12}C(1^+0)}}{\scriptstyle{^{12}C(\bar{\nu},\bar{\nu}')^{12}C(1^+0)}}}= {\epsilon_\nu^2 \over  \pi^2} G_F^2 \sin^4{\theta_W} \cos^2{{\theta \over 2}} e^{-2y} 
\left[-{q_\mu^2 \over 2 q^2} + \tan^2{{\theta \over 2}} \right]  {q^2 \over M^2} \nonumber \\
&&\left[F_1^{(0)} \left({2 \over 3} \psi_{11} + {\sqrt{2} \over 3} \psi_{31}^- + {\sqrt{10} \over 3} \psi_{33} \right) + {\mu^{(0)} \over 2} \left(-{2 \over 3}(1-2y) \psi_{11} -{4 \sqrt{2} \over 3} (1 -{y \over 2}) \psi_{31}^- + {2 \sqrt{10} \over 3} (1-{4 \over 5}y) \psi_{33} \right) \right]^2 
\label{eq:cc12yNC0}
\end{eqnarray}
\end{widetext} 
The isoscalar vector couplings are functions of $q_\mu^2$, with $F_1^{(0)}(0)=1$ and
$\mu^{(0)}(0)=0.88$.

\section{Reduced matrix element functions in Mathematica}
\label{math}

The Mathematica package \emph{SevenOperators} has been developed and made available online \cite{7operatorswebpage} to replace and extend the
tabulated results of Ref. \cite{donnelly}.  The main motivation for
this package is to make calculations of semileptonic electroweak cross
sections and decay rates far less tedious: users can couple it to cross section and rate formulas, explore
various truncations of the density matrix, determine density-matrix
constraints imposed by known rates or cross sections, etc.  Such
closed-form expressions previously would have required tedious hand
calculations.  The \emph{SevenOperators} package has been cross-checked against Ref. \cite{donnelly},
including the production and posting \cite{7operatorswebpage} of tables in the form of Ref. \cite{donnelly} .

In this section we give a basic description of 
\emph{SevenOperators}, and provide useful technical notes.   As an example of usage, we also show its application to the calculation of the  $^{12}$C transitions, which were illustrated in the
previous sections.



The package is available in two forms. 
The first is a notebook version (7operators.nb) that was developed on
Mathematica 5.2.
 The second is a modular version (package.tar.gz)
suitable for use with both the graphical and text-based 
versions of Mathematica. It consists in a .tar.gz archive containing a
master file (7o\_master\_file.m) and several input files, with
self-explanatory names, each of them properly commented.  The whole package is run by simply running the master file.


\emph{SevenOperators} returns seven  functions, which  give the
reduced matrix elements of the seven basic operators between
single-particle states, in the  form of  Eq. (\ref{analytical}). 
The functions have the variables $y,N^\prime,j^\prime, N,j,J$ as
arguments, and their Mathematica syntax is given in Tables
\ref{tablenormal} and \ref{tableabnormal} for normal and abnormal
parity operators respectively.

 \begin{table}[hbt] 
 \caption{\label{tablenormal} The functions describing the reduced matrix elements of the normal parity operators, as implemented in \emph{SevenOperators} Mathematica package \cite{7operatorswebpage}.  For reference, next to each function we give  the reduced matrix element evaluated in \emph{Seven Operators} (see Eqs. (\ref{omegaj})).  Note that the labels $N$ and $j$ fully specify the standard space-spin nonmagnetic quantum numbers for a harmonic oscillator, $n(l~1/2)j$.}
\begin{center}
 \begin{tabular}{|l|l|l|}
\hline
\hline
Mathematica function & Reduced matrix element   \\
\hline
\hline
{\fontfamily{pcr}\selectfont   MJ$[$y,$\{$N$^\prime$,j$^\prime\}$,$\{$N,j$\}$,J$]$}     &  $\langle n^\prime(l^\prime~1/2)j||M^{M_J}_J(\qx)||n(l~1/2)j\rangle$\\
\hline
{\fontfamily{pcr}\selectfont SigmaJ$[$y,$\{$N$^\prime$,j$^\prime\}$,$\{$N,j$\}$,J$]$} & $\langle n^\prime(l^\prime~1/2)j^\prime||\Sigma^{M_J}_J(\qx)||n(l~1/2)j\rangle$ \\

\hline
{\fontfamily{pcr}\selectfont DeltaPJ$[$y,$\{$N$^\prime$,j$^\prime\}$,$\{$N,j$\}$,J$]$} & $\langle n^\prime(l^\prime~1/2)j^\prime||\Delta^{\prime~M_J}_J(\qx)||n(l~1/2)j\rangle$  \\

\hline
\hline
 \end{tabular}
\end{center}
 \end{table}

\begin{table}[hbt] 
 \caption{\label{tableabnormal} The same as Table \ref{tablenormal} for the reduced matrix elements of the abnormal parity operators (see Eqs. (\ref{omegaj})).  Recall that $\Omega^{M_J}_J(\qx)$ and $\Omega^{\prime M_J}_J(\qx)$ are not independent, see Eq. (\ref{omegaprime}). }
\begin{center}
 \begin{tabular}{|l|l|l|}
\hline
\hline
Mathematica function & Reduced matrix element    \\
\hline
\hline
{\fontfamily{pcr}\selectfont DeltaJ$[$y,$\{$N$^\prime$,j$^\prime\}$,$\{$N,j$\}$,J$]$} & $\langle n^\prime (l^\prime~1/2)j^\prime||\Delta^{M_J}_J(\qx)||n(l~1/2)j\rangle$  \\
\hline
{\fontfamily{pcr}\selectfont SigmaPJ$[$y,$\{$N$^\prime$,j$^\prime\}$,$\{$N,j$\}$,J$]$} & $\langle n^\prime (l^\prime~1/2)j^\prime||\Sigma^{\prime~M_J}_J(\qx)||n(l~1/2)j\rangle$  \\
\hline
{\fontfamily{pcr}\selectfont SigmaPPJ$[$y,$\{$N$^\prime$,j$^\prime\}$,$\{$N,j$\}$,J$]$} & $\langle n^\prime (l^\prime~1/2)j^\prime||\Sigma^{\prime\prime~M_J}_J(\qx)||n(l~1/2)j\rangle$ \\
\hline
{\fontfamily{pcr}\selectfont OmegaJ$[$y,$\{$N$^\prime$,j$^\prime\}$,$\{$N,j$\}$,J$]$} & $\langle n^\prime (l^\prime~1/2)j^\prime||\Omega^{M_J}_J(\qx)||n(l~1/2)j\rangle$ \\
\hline
{\fontfamily{pcr}\selectfont OmegaPJ$[$y,$\{$N$^\prime$,j$^\prime\}$,$\{$N,j$\}$,J$]$} & $\langle n^\prime (l^\prime~1/2)j^\prime||\Omega^{\prime M_J}_J(\qx)||n(l~1/2)j\rangle$ \\
\hline
\hline
 \end{tabular}
\end{center}
 \end{table}



Notice that \mt\ has built in functions for the Wigner 3j and 6j symbols.  For the  practical reason of avoiding warning messages in the output, we explicitly imposed that a 3j symbol is put to zero when the conditions for its existence are not satisfied, while the built-in function is used otherwise.  A similar procedure was used for 6j symbols. 

There is no built-in function for the  Wigner 9j symbols in \mt, therefore we had to build one, using the definition:
\begin{equation}
\left\{ \begin{array}{ccc}
j_1  &  j_2 & J_{12} \\
j_3  & j_4  &  J_{34}\\
J_{13} & J_{24}  & J
\end{array} \right\} =  
\sum_{g}  (-1)^{2 g } (2 g +1) 
\left\{ \begin{array}{ccc}
j_1  & j_2 & J_{12} \\
J_{34} & J   & g
\end{array} \right\}
\left\{ \begin{array}{ccc}
 j_3 & j_4 & J_{34} \\
 j_2 & g   & J_{24}
\end{array} \right\}
\left\{ \begin{array}{ccc}
J_{13}  & J_{24} & J \\
g & j_1   & j_3
\end{array} \right\}~. 
\label{9j}
\end{equation}

The code implements an explicit check of the conditions to have
non-zero reduced matrix elements: (i) parity conservation, (ii) that
the nodal numbers $n$ and $n^\prime$ are positive: $n>0$,
$n^\prime>0$, and (iii) that $j,j^\prime,J$ satisfy the triangular
inequality. Only if all three conditions are satisfied, the code
proceeds with the evaluation of the function describing the reduced
matrix element, otherwise the function is put to zero.

In Appendix A we include the text of a Mathematica notebook (also available
online \cite{7operatorswebpage}) where \emph{SevenOperators} used
to calculate the first of the three examples we considered,
$^{12}$C$(\nu_e,e^-)^{12}$N(g.s.), yielding the result shown in
Eq. (\ref{eq:cc12y}).  


 We acknowledge support from the U.S. Department of Energy under grants DE-FG02-00ER-41132
 and DE-FC02-01ER41187 (SciDAC), and from ORNL contract 4000058152.


\section*{Appendix A: Mathematica notebook calculation of $^{12}$C($\nu_e,e^-)^{12}$N($1^+1$)}
\label{mathapp}
This notebook illustrates the uploading of the package
(specifically, uploading the master file of the
modular version) and the use of the seven basic operator matrix elements
to construct the nuclear multipole operators in Sec. \ref{relationship}.  The application is
the charge-current cross section $^{12}$C($\nu_e,e^-)^{12}$N($1^+1$), the first example
discussed in this paper.  The notebook describes points where a user will need to
insert steps to account for the specific electroweak reaction, assumed one-body density
matrix, isospin matrix elements, and allowed multipolarities.
\emph{SevenOperators} makes
an otherwise tedious computation very immediate, eliminating much of the potential for
algebraic errors.  The final result in this notebook can be further manipulated to improve
its form, but we have not done this here, in order to keep this appendix readable.
Such final manipulations are also a matter of personal taste.

\pagebreak

\begin{figure}[htbp]
  \centering
   \includegraphics[width=18cm]{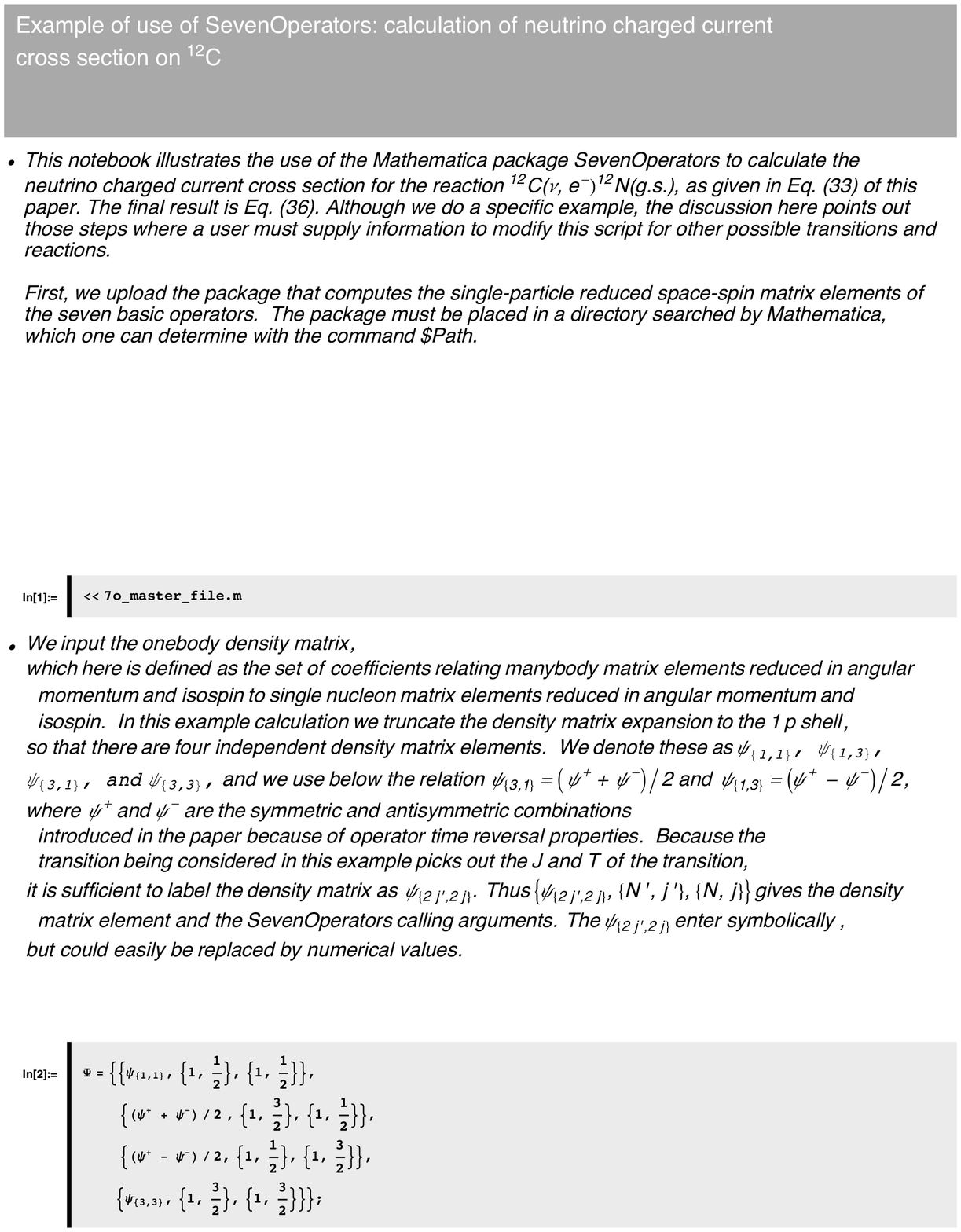}
\end{figure}

\begin{figure}[htbp]
  \centering
   \includegraphics[width=18cm]{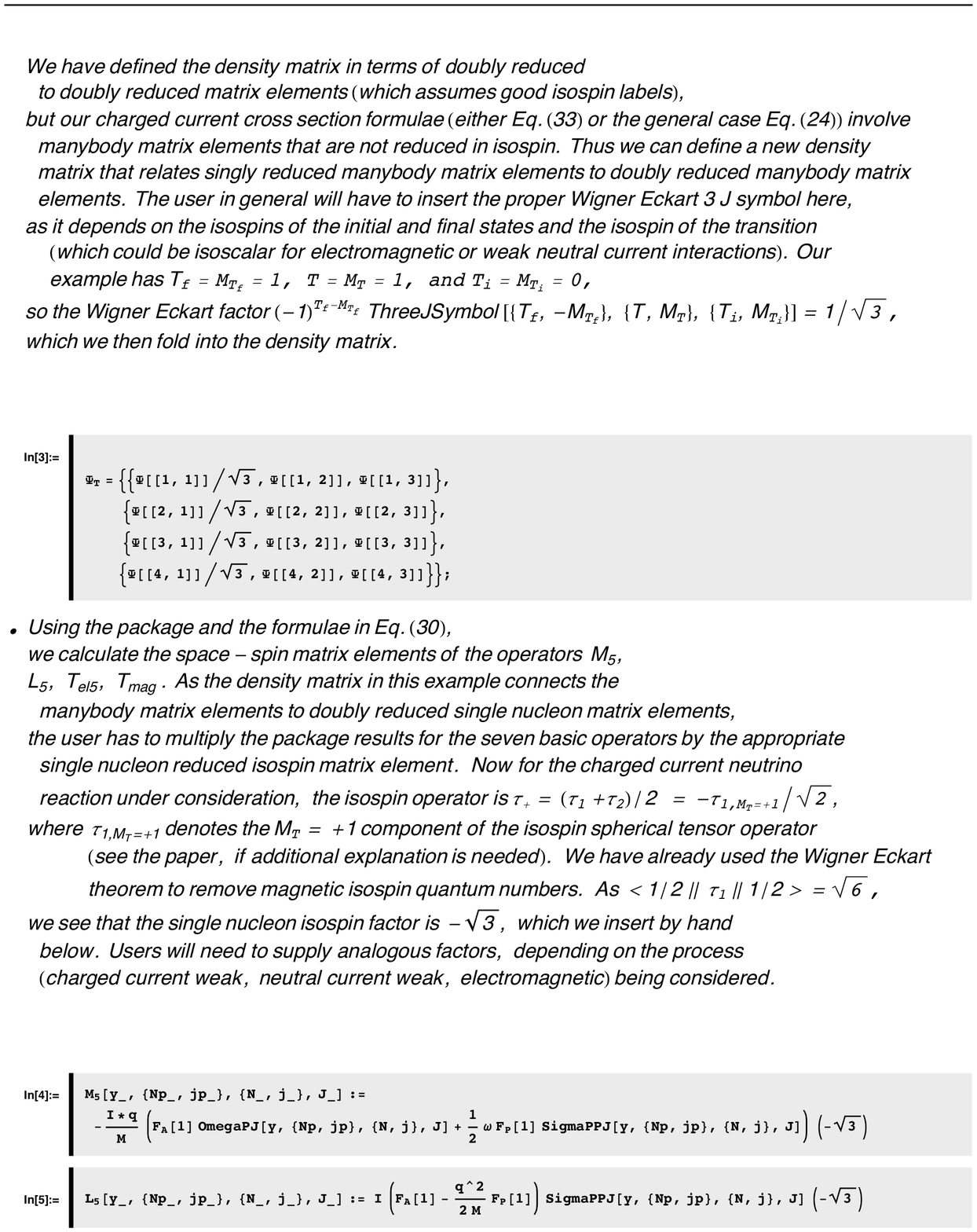}
\end{figure}

\begin{figure}[htbp]
  \centering
   \includegraphics[width=18cm]{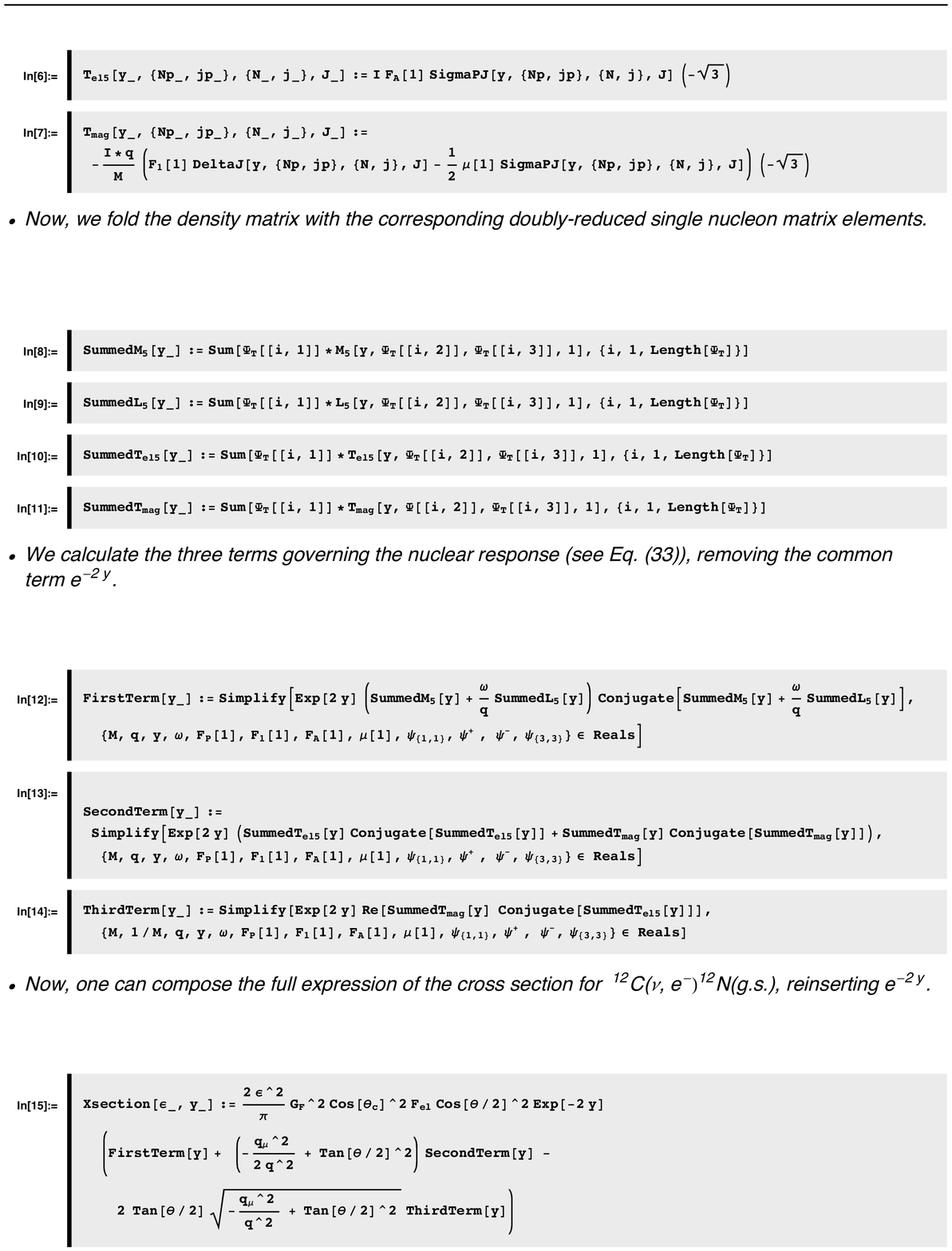}
\end{figure}

\begin{figure}[htbp]
  \centering
   \includegraphics[width=18cm]{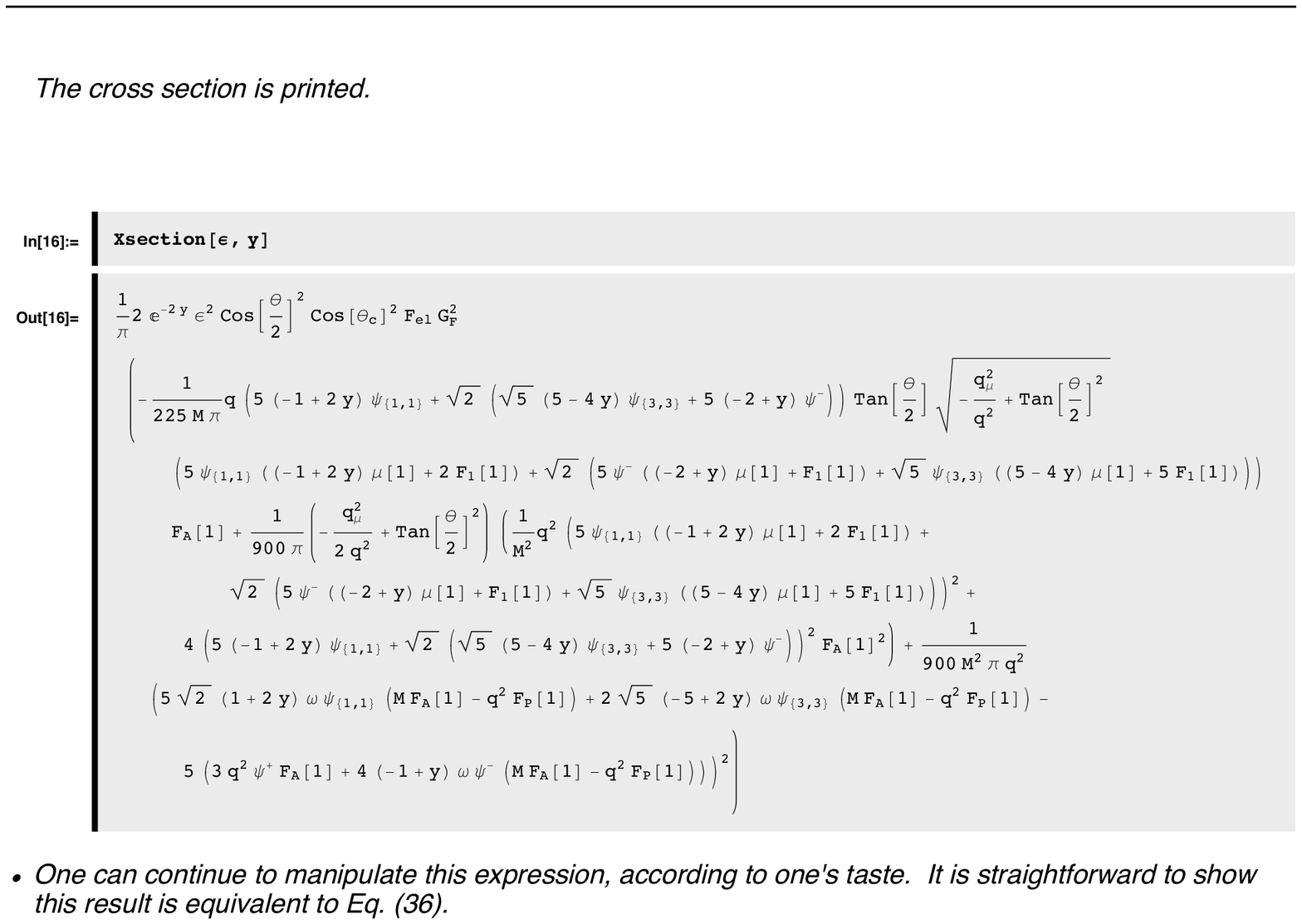}
\end{figure}


\end{document}